\documentclass[prl, reprint,showpacs]{revtex4-1} 
\usepackage{amsmath,amssymb,amsfonts, graphics,graphicx,color,multirow}

\begin{document}
\title{Efficiency of surface-driven motion: nano-swimmers beat micro-swimmers\\}
\author{Benedikt \surname{Sabass}}
\author{Udo \surname{Seifert}} 
\affiliation{II. Institut f\"ur Theoretische Physik, Universit\"at Stuttgart,
70550 Stuttgart, Germany}
\begin{abstract}
Surface interactions provide a class of mechanisms which can be employed for
propulsion of micro- and nanometer sized particles. We investigate the related
efficiency of externally and self-propelled swimmers. A general scaling relation
is derived showing that only swimmers whose size is comparable to, or smaller
than, the interaction range can have appreciable efficiency. An upper bound for
efficiency at maximum power is 1/2. Numerical calculations for the case of
diffusiophoresis are found to be in good agreement with analytical expressions
for the efficiency. 
\end{abstract}
\pacs{47.57.J-, 82.70.-y, 47.63.mf}
\maketitle
\paragraph{Introduction.--}In recent years, much effort has been spent to
understand biological micro-swimmers and to develop their artificial
counterparts \cite{lauga2009hydrodynamics}. Mechanisms for propulsion of
micron-scale objects include, e.g., flagellar motors, surface streaming and
non-reciprocal shape distortions. A different class of mechanisms are based on
surface interactions which convert gradients in the environment of the swimmer
into a hydrodynamic flow around the particle, thus propelling it forward. A
paradigmatic example is diffusiophoretic motion in chemical gradients
\cite{anderson1989colloid}. Here, theoretical and experimental advances have
lately furnished a good understanding
\cite{golestanian2005propulsion,dhar2006autonomously, howse2007self,
ruckner2007chemically, popescu2009confinement, julicher2009generic,
Bocquet_PRL}. Miscellaneous phoretic mechanisms are based, e.g., on gradients in
temperature or electrical fields. In contrast to other ways of self-propulsion,
they require no mechanical deformation of the swimmer. They are thus very
expedient to use for synthetic nanomotors whose mechanical degrees of freedom
are hard to control. In contrast to the efficiency of molecular motors
\cite{Julicher_motor} or other swimming mechanisms at low Reynolds numbers
\cite{stone1996propulsion, avron2004optimal, teran2010viscoelastic}, the
efficiency of surface-driven propulsion has, to our knowledge, received no
attention so far. However, this question becomes relevant when energy resources
are limited by the environment of the swimmer. Furthermore, if the envisioned
application dictates a high swimming velocity, the quadratic speed dependence of
the dissipation will bring energetic aspects to attention. Industrial
applications of diffusiophoresis \cite{anderson1984diffusiophoresis} employ a
high number of particles and therefore efficiency may become rather important
here. Likewise, large dissipation could lead to undesired side effects like
local heating. One might also ask from a biological perspective for the
advantage of motility based on active processes on or near the surface
\cite{lammert1996ion, blake1971spherical}. These issues motivate us to
investigate generic features of the efficiency of surface-driven motion, both,
for externally and self-propelled swimmers.
\paragraph{Model.--}The swimmer, a spherical particle with radius $R$, is
surrounded by a multicomponent fluid in a large container. The translational
velocity of the swimmer in the laboratory frame is $\mathbf{V}$ and it does not
rotate. A constant, externally applied, force $\mathbf{F}_{\rm m}$ may act on
the particle. The Reynolds number is assumed to be small enough that the fluid
can be described by the Stokes equation. The variety of mechanisms which can
generate hydrodynamic flow near the surface employ very different sources of
energy and differ accordingly in their thermodynamic description. In an
isothermal steady state the overall energy input is given by the entropy
production of the system multiplied by temperature plus the external work
delivered by the swimmer. These quantities can in principle be calculated in the
framework of irreversible thermodynamics, thus allowing for the computation of
efficiencies. Yet, the results are usually not analytically accessible because
nonlinear field equations must be solved. Seeking a more general answer to the
question of efficiency of surface-driven processes we here leave
process-specific details aside. Instead, we concentrate on the hydrodynamic
efficiency $\epsilon_{\rm h}$ which is a common upper bound for the true
efficiency $\epsilon$. The (positive) power output is given by $P_{\rm o} \equiv
-\mathbf{V}\mathbf{F}_{\rm m}$. The (positive) hydrodynamic power input $P_{\rm
h,i}$ is then defined as the sum of power output and hydrodynamic dissipation
leading to a bound on the true efficiency 
\begin{equation}
\epsilon \leq \epsilon_{\rm h} \equiv \frac{P_{\rm o}}{P_{\rm h,i}} =
\frac{-\mathbf{V}\mathbf{F}_{\rm m}}{-\mathbf{V}\mathbf{F}_{\rm m}+2 \eta\,\int
\mathbf{E}:\nabla\mathbf{v}\,\mathrm{d}V}
\label{efficiency}
\end{equation}
where $\mathbf{v}$ represents the velocity and $\eta$ the viscosity of the
fluid. $\mathbf{E} = \left(\nabla \mathbf{v}+(\nabla \mathbf{v})^{T}\right)/2$
is the deviatoric strain rate. We employ the customary assumption that $\nabla
\cdot \mathbf{v} = 0$ in the limit of dilute solutes. This implies, besides
incompressibility, that the mass densities of all fluid components are similar.
The efficiency Eq. (\ref{efficiency}) depends on the external force
$\mathbf{F}_{\rm m}$. However, in order to find a typical value of
$\epsilon_{\rm h}$, we focus on the hydrodynamic efficiency at maximum power
output $\epsilon^{*}_{\rm h}$, which eliminates $\mathbf{F}_{\rm m}$.
\paragraph{Scaling of efficiency of micro-swimmers.--}The lengthscale of the
surface interactions is typically on the order of several $\rm nm$. For particle
sizes on the order of $\mu \rm m$ it therefore makes sense to split the
hydrodynamic problem into an inner problem where the fluid speed is strongly
influenced by the surface interaction and an outer problem where the direct
influence of the interaction is negligible. This is the classical boundary layer
approximation which we will use in the following. 
We start with an estimate for power output $P_{\rm o}$. The external force in
our expression for power output $-\mathbf{F}_{\text{m}} \mathbf{V}$ demands for
the presence of a long ranged stokeslet \cite{happel1983martinus} and $P_{\rm o}
\sim \eta R |\mathbf{V}|^2$. Next, we estimate the power input $P_{\rm h,i}$
consisting of the hydrodynamic work rate in the inner and the outer region. The
work rate in the outer region can be calculated from ordinary hydrodynamics and
scales as $ \eta R |\mathbf{V}|^2$. The dissipation in the boundary layer
deserves a slightly more careful analysis. Here the thinness of the layer
together with a no-slip condition at the surface of the particle leads to a
drastic change of fluid velocity. This implies strong viscous dissipation. The
thickness of the hydrodynamic boundary layer is denoted by $L$. The radial
derivative is $\sim 1/L$ and the speed normal to the surface is small due to the
impermeable boundary. Hence, the dissipation rate per volume in the boundary
layer can be approximated as $2 \eta\, \mathbf{E}:\nabla \mathbf{v} \sim \eta
|\mathbf{V}|^2/ L^2$. When performing the volume integral over the local
dissipation rate, the dominant contribution comes from a volume near the surface
which we approximate with $R^2 L$. Taken together, we find for the power input
$P_{\rm h,i} \sim \eta |\mathbf{V}|^2 R^2/ L$ where we have already dropped
$P_{\rm o}$ and the dissipation outside the boundary region because their
relative contribution is $O(L/R)$. Taken together, we find that the hydrodynamic
efficiency scales to leading order in $L/R$ as
\begin{equation}
\epsilon_{\rm h} \sim L/R.
\label{eq_eps_scaling}
\end{equation}
This generic scaling shows that any surface interaction whose range is
considerably smaller than the size of the particle is inefficient in driving
it. 
\paragraph{Upper bound on efficiency.--}The above estimate for the hydrodynamic
efficiency loses its validity for nano-swimmers where $L$ is comparable to $R$.
To derive a general upper limit for the hydrodynamic dissipation we compare the
dissipation rate of the true fluid velocity field $\mathbf{v}$ with the
dissipation rate in an auxiliary velocity field $\mathbf{v}'$ around a passively
dragged particle. The true fluid velocity $\mathbf{v}$ is driven by any velocity
independent body force $\mathbf{b}$, arising, e.g., from surface interactions
and hence $\eta \nabla^2 \mathbf{v} - \nabla p= -\mathbf{b}$. $\mathbf{v}'$
satisfies the homogeneous Stokes equation $\eta \nabla^2 \mathbf{v}' - \nabla
p'=0$ and $\nabla \cdot \mathbf{v}'=0$. The boundary conditions for
$\mathbf{v}'$ are to be the same as for $\mathbf{v}$. Starting with the
inequality
$\left(\mathbf{E}'-\mathbf{E}\right):\left(\mathbf{E}'-\mathbf{E}\right) \geq 0$
one finds
\begin{equation}
2 \eta\,\int \mathbf{E}:\nabla\mathbf{v}\,\mathrm{d}V \geq \int 
\nabla\cdot\left(\mathbf{v}  \left[-p'\mathbf{I}+2\eta
\mathbf{E'}\right]\right)\,\mathrm{d}V = \mathbf{V}\mathbf{T} \mathbf{V}
\label{eq_diss_bound}
\end{equation}
where $\mathbf{T}$ is the resistance tensor of translation, which is for a
spherical particle given by $6 \pi \eta R\, \mathbf{I}$. Also, due to linearity
of the Stokes equation the power output can be written as ~$P_{\rm o} =
-\mathbf{V}\,\left[\mathbf{T}\mathbf{V} + \mathcal{F}\right]$ where
$\mathcal{F}$ is a function of the surface interaction forces but independent of
swimming speed. Explicit expressions for $\mathcal{F}$ can be obtained \cite{teubner1982motion}, but are
not required here.
The particle velocity at maximum power output is ~$\mathbf{V}^{*}=
-\mathbf{T}^{-1}\mathcal{F}/2 $. Using $\mathbf{V}^{*}$ and
Eq.~(\ref{eq_diss_bound}) in Eq.~(\ref{efficiency}) yields
\begin{equation}
\epsilon^{*}_{\rm h} \leq 1/2.
\end{equation}
Remarkably, this is in a quite general sense an upper bound for the efficiency
at maximum power of any hydrodynamic motor in the Stokes regime. It is formally
related to results for heat engines \cite{VanDenBroeck_PRL2005} but differs in
the definition of efficiency and in that we are dealing with hydrodynamic
systems. As demonstrated by the numerical calculations below, the upper limit
for hydrodynamic efficiency can be almost achieved by small swimmers. In the
remainder of this letter we shall support these general considerations by a
detailed treatment of diffusiophoresis where interactions with gradients of
ionic or neutral solutes drive the particle.
\paragraph{Diffusiophoresis.--}As customary, the system is treated in the
quasi-stationary limit. We employ a spherical coordinate system aligned in the
$\mathbf{\hat{e}}_3$ direction where $r$ is the distance from the particle
center and $\theta$ is the inclination angle in the axisymmetric problem. The
potential $\Psi(r, \theta)$ mediates interactions between the swimmer and the
solute concentration fields $c_i(r,\theta)$ in the dilute limit. In the case of
ionic solutes, symmetrically charged cations ($c_1$) and anions ($c_2$) with
different mobilities are present. $\Psi$ must then be determined from Poisson's
equation $ \nabla^2\Psi = -4\pi (Z e)^2 \varepsilon^{-1} \left(c_1-c_2\right)$
where $Ze$ is the charge of each ion and $\varepsilon$ is the dielectric
constant of the fluid. The range of the ionic potential is determined by the
Debye length $l =\kappa^{-1} \equiv \left[8 \pi (Z e)^2
c^{\infty}(0)/(\varepsilon kT)\right]^{1/2}$ where $c^{\infty}(0)$ is the
concentration at $r=0$ in absence of a swimmer \cite{keh2000diffusiophoretic}.
$kT$ is the thermal energy scale. For the case of a non-ionic concentration
gradient we use only one kind of solute ($c_1$) interacting with the swimmer via
an arbitrary, radially symmetric potential $\Psi(r)$ which also decays on some
lengthscale $l$. The resulting steady state solute fluxes are $\mathbf{j}_{1}$,
$\mathbf{j}_{2}$ in the ionic, and  $\mathbf{j}_{1}$ in the non-ionic case,
respectively. When neglecting convection, we have for solute conservation
\begin{align}
0=\nabla\cdot \mathbf{j}_{1,2}  = \nabla\cdot\left[-D_{1,2} \left(\nabla c_{1,2}
\pm (kT)^{-1}\,c_{1,2}\nabla \Psi  \right)\right]
\label{eq_diff}
\end{align}
where $D_{1}$, $D_{2}$ are the diffusion constants of the solutes. For
diffusiophoresis of a passive swimmer in an externally maintained concentration
gradient the boundary conditions are ~$\mathbf{\hat{e}}_{\rm
r}\,\mathbf{j}_i|_{r=R} = 0$ and ~$\left(\nabla c_i \right)|_{r\rightarrow
\infty} = \mathrm{const.} \times \mathbf{\hat{e}}_3$. Also, the boundary
conditions of an ionic potential are determined such that the electric current
vanishes at infinity $\left(\mathbf{j}_1 -\mathbf{j}_2\right)|_{r\rightarrow
\infty} = 0$ \cite{prieveElectrolytes}. Both, the ionic and non-ionic solutes
mediate a body force $\mathbf{b}$ given by $-\left(c_1-c_2\right)\nabla \Psi$
and $-c_1\nabla \Psi$, respectively. Accordingly, the Stokes equation with $\pm
\mathbf{F}_{\rm m}\parallel\mathbf{\hat{e}}_3$ becomes
 \begin{align}
\eta \nabla^2 \mathbf{v} - \nabla p =\mathbf{-b},& &\mathbf{v}|_{r=R} =0, & &
\mathbf{v}|_{r\rightarrow \infty} =  -V \mathbf{\hat{e}}_3.
\label{eq_stok}
\end{align}
This model is only valid if the mutual interactions of solutes with radius $a$
are negligible. The corresponding corrections to the diffusion coefficient are
proportional to the volume fraction and we hence demand $c\,a^3 \ll 1 $.
Moreover, the relative corrections of the solute-swimmer interactions are $\sim
(a/R)^2$, which should also remain $\ll 1$ if the solute size is to be
neglected.
This restricts our model to swimmers which are at least one or two magnitudes
larger than the solutes. For solutes in the \AA{} range, we hence require
$R\,\gtrsim30\rm\, nm$. Then, the nano-swimmer regime corresponds to a
''diffusiophoretic Debye-H\"uckel limit''.
\paragraph{Diffusiophoretic efficiency.--}In order to explicitly confirm the
scaling of micro-swimmer efficiency, Eq.~(\ref{eq_eps_scaling}), we employ the
established theory for diffusiophoresis when $l \ll R$
\cite{anderson1989colloid, prieveElectrolytes}. The smallness of $l$ implies
that the normal concentration profile near the surface is near equilibrium. Then
$c_{1,2}(y,\theta) \approx  \tilde{c}(\theta) \exp \left[\mp\Psi(y)/kT\right]$
where $y\equiv r-R$ and $\tilde{c}(\theta)$ is the undisturbed concentration of
solutes outside the boundary layer. To leading order in $l/R$ the radial body
force is compensated by a radial pressure change $\mathbf{\hat{e}}_r\mathbf{b}
\approx \partial_y p$ which yields $p\approx p_0 + kT \sum_i \left[c_i(y,\theta)
-\tilde{c}(\theta)\right]$. The leading order contribution of the Stokes
equation for lateral flow is $\eta \partial_{y}^2 v_{\theta} -
R^{-1}\partial_\theta p \approx -\mathbf{\hat{e}}_{\theta}\mathbf{b}$. Upon
insertion of the pressure and integration one obtains the boundary layer fluid
velocity in a comoving frame \cite{anderson1989colloid}
\begin{equation}
 v_{\theta}(y,\theta) \approx -\frac{kT}{\eta}\int_0^y\int_{y'}^{\infty}f(y'')
dy'' dy'\, \frac{\partial_\theta \tilde{c}(\theta)}{R}
 \label{eq_vsurf}
\end{equation}
with the function $f(y)$ given below \cite{fy_note}. Extension of the integral
limit to infinity yields the so-called slip velocity $v_{\rm s}(\theta) \equiv
v_{\theta}(y,\theta)|_{y\rightarrow \infty}$ at the interface between boundary
layer and the outer flow. To calculate the particle speed $V$ from the slip
velocity one matches $v_{\rm s}$ as boundary condition to an outer hydrodynamic
solution where $\mathbf{b} = 0$. The result is $V \approx \frac{F_{\rm m}}{6 \pi
\eta R} + \frac{1}{2 }  \int_0^{\pi} v_{\rm s}(\theta) \sin^2 \theta \,
\mathrm{d} \theta$. The efficiency is determined by inserting the inner solution
Eq.~(\ref{eq_vsurf}) into Eq.~(\ref{efficiency}) while the matched outer
solution does not contribute in leading order. The result for the efficiency at
maximum power becomes
\begin{equation}
\epsilon^{*}_{\rm h} \approx  \frac{9 \left(\int_0^{\pi} \partial_{\theta}
\tilde{c}(\theta) \sin^2\theta \, \mathrm{d}\theta\right)^2}{32 \int_0^{\pi}
\left(\partial_{\theta} \tilde{c}(\theta)\right)^2\sin\theta\, \mathrm{d}
\theta} \frac{L_{\rm D}}{R}
\label{eff}
\end{equation}
where we have defined the dissipation length
\begin{equation}
L_{\rm D} \equiv \frac{\left[\int_0^{\infty}y f(y)
\mathrm{d}y\right]^2}{3\,\int_0^{\infty}f(y)\int_0^{y}y' f(y')\mathrm{d}y'
\mathrm{d}y}
\label{Lss_formula}.
\end{equation}
The dissipation length $L_{\rm D}$ is a measure for the radial extension of the
layer where velocity gradients are strong. It is expected to be similar to the
thickness of the layer where the fluid is driven by the body forces. For ionic
solutes $L_{\rm D}$ is to leading order proportional to the Debye length
$\kappa^{-1}$ \cite{LDionic_note}.\\ 
In the case of non-ionic solutes the expression for $L_{\rm D}$ depends on the
choice of $\Psi(r)$. Here it is of interest to compare $L_{\rm D}$ with the
thickness of the layer of excess solute $L_{\rm \Psi}$
\cite{Ls_note,anderson1989colloid} because the latter can be inferred indirectly
by measuring the diffusiophoretic speed. It is usually on the order of $10\,\rm
nm$ \cite{staffeld1989diffusion}. We find $L_{\rm D} = L_{\rm \Psi}= l/2$ in the
concrete cases of a hard-core repulsion as well as for $\Psi(y)/kT = (l/y)^n$
with $n\rightarrow \infty$ and small $l$. Moreover, if one replaces the inner
integral limit $y$ in the denominator of Eq.~(\ref{Lss_formula}) with $\infty$
then $L_{\rm D} = L_{\Psi}/3$. This, together with evaluations of
Eq.~(\ref{Lss_formula}) for various $\Psi(r)$ shows that it is safe to estimate
the magnitude of $L_{\rm D}$ through $|L_{\Psi}|$ in the non-ionic case. 

We have also investigated the effect of hydrodynamic slip boundary conditions
\cite{ajdari2006giant} on $\epsilon^{*}_{\rm h}$. These reduce the hydrodynamic
dissipation in the boundary layer but additional dissipation occurs directly at
the surface. Effectively, $L_{\rm D}$ is increased by a few $\rm nm$. A good
absolute value for efficiency can, however, not be achieved in this way when
$L_{\rm D} \ll R$ is still valid. 
\paragraph{Numerical analysis.--}
In order to extend our analysis to nanoparticles where $l \gtrsim R$ we solve
the coupled diffusion and hydrodynamic equations numerically. In the case of
ionic solutes the nonlinearities are avoided by expanding the solution for low
dimensionless surface charge density $q \equiv 4\pi Ze Q (\varepsilon \,\kappa
kT)^{-1}$ as done in \cite{keh2000diffusiophoretic}. The constant $q$ can then
be related to the electrostatic potential on the surface $\Psi(R) =
kT\,q\,\kappa R\left(1+\kappa R\right)^{-1}\,+O(q^3)$. For non-ionic solutes we
can directly solve Eqs.~(\ref{eq_diff}) and (\ref{eq_stok}) after choosing a
potential $\Psi(r)$.

\begin{figure}[ht]
\begin{center}
 \includegraphics[scale=0.63]{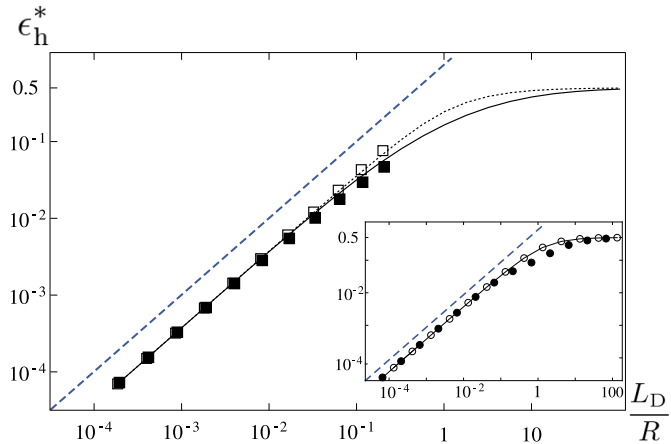}
\end{center} 
\caption{Efficiency at maximum power $\epsilon^{*}_{\rm h}$ for diffusiophoresis
in non-ionic solutes. ($\square$): Van der Waals attraction $\Psi(y) = -16/9 \,A
 \,a^3R^3 y^{-3} (y+2R)^{-3}$ where $A=1\,kT$ and $a$ is the solute radius. $a/R
=\left[10^{-4}\ldots 0.1\right]$ and attraction is truncated at $y=a$.
Concentration gradient is established externally. ($\blacksquare$): Same
$\Psi(y)$ as before but concentration gradient is produced by a Janus particle.
($\cdots\cdots$): Generic repulsion $\Psi(y)=kT\,\exp\left(-y/l\right)$ with
$l/R =\left[10^{-4}\ldots 10^{2}\right]$. Concentration gradient is established
externally. (\textemdash\textemdash): ~Same $\Psi(y)$ as before but
concentration gradient is produced by a Janus particle. Inset: diffusiophoresis
in an externally established gradient of ionic solutes and $\kappa^{-1}/R
=\left[10^{-4}\ldots 10^{2}\right]$. ($\bullet\,\bullet\,\bullet$): ~$q=0.1,\,
\nu =0$. ($\circ\,\circ\,\circ$): ~$ q=-0.5,\,\nu =10$.
(\textemdash\textemdash): ~$q=0.1, \, \nu =1$.}
\label{fig_2}
\end{figure}

The resulting efficiency at maximum power $\epsilon_{\rm h}^{*}$ is displayed in
Fig.~\ref{fig_2}. Note that $\epsilon^{*}_{\rm h}$ is independent of $\eta$ and
$\left(\nabla c_i \right)|_{r\rightarrow \infty}$. The absolute concentration
level enters $\epsilon^{*}_{\rm h}$ only for ionic solutes via the Debye length.
For $L_{\rm D} \lesssim R$ we find good agreement of the numerical results with
the boundary layer theory where Eq.~(\ref{eff}) predicts $\epsilon^{*}_{\rm
h}=(3/8) \,L_{\rm D}$ since the concentrations $c_{1,2}(r,\theta)$ are linear in
$\cos\theta$. For $L_{\rm D} > R$ the efficiencies of nano-swimmers are actually
in the range of the upper bound given by $1/2$. This is due to the low
dissipation when the swimmer moves in a spatially slowly varying 
flow field. The deviations from the analytical predictions are most pronounced
in the intermediate regime of $L_{\rm D} \simeq R$. For ionic solutes (inset of
Fig.~\ref{fig_2}) we find that the efficency is larger when the two solutes have
different mobilities $\nu  \neq 0$; see \cite{fy_note}. Then electrophoresis in
self-generated concentration disturbances around the particle influences the
swimming.
\paragraph{Janus particles.--}To complement the results for passive swimmers we
also investigate active Janus particles where the neutral solute concentration
gradient is maintained by a chemical reaction on the surface
\cite{golestanian2005propulsion,dhar2006autonomously, howse2007self,
ruckner2007chemically, popescu2009confinement}. This is modeled through the
boundary conditions of Eq.~(\ref{eq_diff}) by  ~$\mathbf{\hat{e}}_{\rm
r}\,\mathbf{j}_1|_{r=R} = \alpha(1+\cos\theta)$ and ~$ c_1|_{r\rightarrow
\infty} = \mathrm{const.}$ where $\alpha$ is the effective solute production
rate per area. The efficiency $\epsilon^{*}_{\rm h}$ becomes independent of
$\alpha$ and differs only in the regime of $L_{\rm D} \simeq R$ from the results
for swimmers in an externally applied gradient. Active Janus particles are in
this regime hydrodynamically less efficient (see Fig.~\ref{fig_2}). This relates
to the fact that long range potentials are not so effective in driving the fluid
when the concentration gradient decays to zero away from the particle. A problem
occurring for small Janus particles is the fast concentration field
homogenization through their rotational diffusion on the timescale of $4 \pi
\eta R^3 (kT)^{-1}$. It may thus be necessary to fix such motors directionally.
\\
We also mention that the full efficiency $\epsilon$ of Janus particles has to
take into account losses due to maintenance of spatial concentration gradients
and due to chemical reactions. The power input then reads $P_{\rm i}=P_{\rm
i,h}- \int \left(\sum_i \mathbf{j}_i\nabla\mu_i+ \sum_{k}A_k\, r_{k}  \right)
\mathrm{d}V$ where $\mu_i$ is a chemical potential of species $i$ including
$\Psi_i$. $A_k$ and $r_{k}$ are affinity and rate of the $k$-th chemical
reaction. 
\paragraph{Swimming speed.--}The energetic differences of micro- and
nano-swimmers are accompanied by differences in the swimming speed. As a simple
example, for micro-swimmers in neutral solute with $\mathbf{F}_{\rm m} =0$ one
has $V \sim  \frac{kT}{\eta}\, l^2 \left(\nabla
c_1\right)|_{r\rightarrow\infty}$ \cite{anderson1989colloid}. However for
swimmers with $l \gg R$ we find the scaling $V \sim  \frac{kT}{\eta}\,
\frac{l^3}{R} \left(\nabla c_1\right)|_{r\rightarrow\infty}$ \cite{Sabass2010}.
Hence, aside from efficiency, nano-swimmers are qualitatively different from
micro-swimmers in that their size can matter for their mobility. 
\paragraph{Conclusion.--}Although phoretic effects are known for more than a
century their energetic aspects have hardly been explored. In this letter we
make an attempt in this direction by suggesting a generic scaling relation for
the efficiency of surface-driven motion. It provides a widely applicable and
simple concept to estimate hydrodynamic efficency without detailed knowledge of
the system. Further, we show with analytical and numerical calculations that
phoretic nano-swimmers offer energetic advantages; in particular with ionic solutes, where the Debye
length can be tuned. Taken together, we see inspiring perspectives for artificial
nanomotors which, reminiscent of actual biological motors, could possibly move
not only in a controllable but also in an efficient way. 
%

\end{document}